\documentclass[%
superscriptaddress,
amsmath,
amssymb,
aps,
prb,
twocolumn,
]{revtex4-2}

\usepackage{graphicx}
\usepackage{dcolumn}
\usepackage{bm}
\usepackage{ulem}
\usepackage{times}
\usepackage{bbding}

\usepackage[colorlinks,linkcolor=blue,citecolor=blue,urlcolor=blue]{hyperref}

\def\bs#1{\boldsymbol{#1}}

\begin{document}


\title{Geometrical Nonlinear Hall Effect Induced by Lorentz Force}

\author{Junjie \surname{Yao}}
\affiliation{State Key Laboratory of Low Dimensional Quantum Physics and Department of Physics, Tsinghua University, Beijing, 100084, China}

\author{Yizhou \surname{Liu}}
\email{yizhouliu@tongji.edu.cn}
\affiliation{School of Physics Science and Engineering, Tongji University, Shanghai 200092, China}
\affiliation{Shanghai Key Laboratory of Special Artificial Microstructure Materials and Technology, Tongji University, Shanghai 200092, China}

\author{Wenhui \surname{Duan}}
\affiliation{State Key Laboratory of Low Dimensional Quantum Physics and Department of Physics, Tsinghua University, Beijing, 100084, China}
\affiliation{Collaborative Innovation Center of Quantum Matter, Beijing 100084, China}
\affiliation{Institute for Advanced Study, Tsinghua University, Beijing 100084, China}


\date{\today}

\begin{abstract}
The recently discovered nonlinear Hall (NLH) effect arises either without external magnetic field (type-I) or with an in-plane magnetic field (type-II). In this work we propose a new type of geometrical nonlinear Hall effect with an out-of-plane magnetic field (type-III) induced by the combination of Lorentz force and anomalous electronic velocity. The type-III NLH effect is proportional to the more refined structures of Bloch wave functions, i.e., the dipole moment of square of Berry curvature, thus becoming prominent near the band crossings or anticrossings. Our effective model analysis and first-principles calculations show that gate-tuned MnBi$_2$Te$_4$ thin film under uniaxial strain is an ideal platform to observe this effect. Especially, giant unidirectional magnetoresistance can occur in this material, based on which an efficient electrical transistor device prototype can be built. Finally a symmetry analysis indicates that type-III NLH effect has unique symmetry properties stemming from Berry curvature square dipole, which is different from other previously reported NLH effects and can exist in a wider class of magnetic crystals. Our study offers new paradigms for nonlinear electronics.
\end{abstract}

\maketitle

\section{Introduction} The Hall effects, as one of the most long-history and yet paradigmatic phenomena in condensed matter physics, have been extensively studied due to their underlying rich physics \cite{Popovic2003book, Cage2012book, Chien2013book}. Interestingly, the Hall effects can have two very different physical origins: they can be induced by the Lorentz force in the external magnetic field, or result from the electronic band geometry and topology \cite{Nagaosa2010May, Xiao2010Jul}. Prominent examples of the latter include the quantum Hall effect in a strong magnetic field \cite{Klitzing1980Aug, Thouless1982Aug} and its anomalous version without a magnetic field \cite{Haldane1988Oct, Yu2010Jul, Chang2013Apr}, which have provided valuable insights into the nontrivial electronic structures related to the momentum space textures (i.e., curvature and metric) of Bloch wave functions, and have triggered exotic applications in many other systems \cite{Xiao2006Jul, Son2013Sep, Xu2014Jun, Nandy2017Oct}. The former, at lower magnetic fields, is classical and does not rely on Bloch wave function textures \cite{Nagaosa2010May}.

\begin{figure*}
    \centering
    \includegraphics[width=\linewidth]{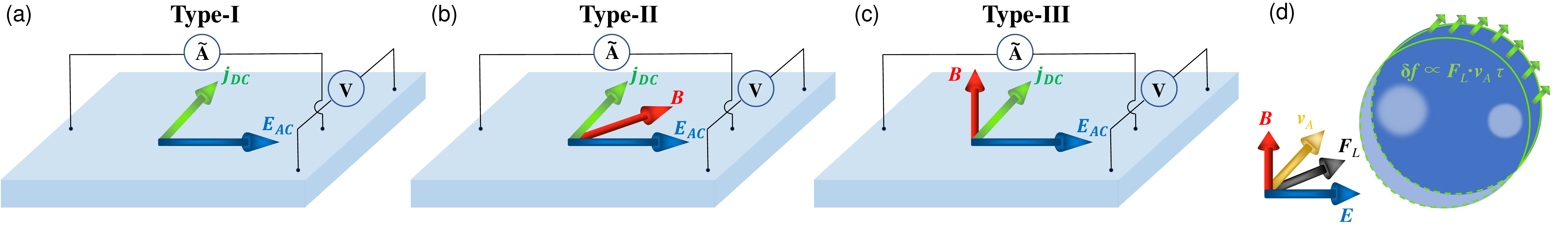}
    \caption{\label{fig1}
    (a)-(c) Schematic setups of type-I, type-II, and type-III NLH effects, respectively. In the NLH effects, a transverse dc signal can be generated through a longitudinal ac input. (d) Fermi surface shift induced by the Lorentz force $\bs{F}_L$ and anomalous velocity $\bs{v}_A$ in the type-III NLH effect.
    }
\end{figure*}


On the other hand, the recently discovered nonlinear anomalous Hall (NLAH) effect has stimulated the interest of nonlinear electronic transport and optoelectronic studies. This effect connects the nonlinear response coefficients with some geometrical or topological properties of Bloch wave functions, and has promising applications, including second harmonic generation, radio-frequency ac-dc rectification, and terahertz detection, among others \cite{Du2021Nov, Ortix2021Sep, Gao2014Apr, Sodemann2015Nov, Ma2019Jan, Kang2019Apr, Du2018Dec, Du2019Jul, Tiwari2021Apr, He2021Dec, Du2021Aug, Lai2021Aug, Wang2021Dec, Liu2021Dec, Zhang2021May, Cao2022May, Duan2022Oct, Gao2023Jun, Wang2023Jul, Wang2023Sep, Zhao2023Nov, Kaplan2024, Li2024Jan, Qin2024Jan, Tokura2018Sep, Rikken2001Nov, Rikken2005Jan, Avci2015, Yasuda2016Sep, Fan2019, Guillet2020Jan, Liu2021Feb, Liu2021Oct, Liu2024Jan, Liu2023Mar}. The NLAH effect refers to the Hall current $j^{\textrm{H}} \propto E^2$ in response to the driving electrical field $E$ without external magnetic field as shown in Fig. \ref{fig1} (a), which we dub type-I nonlinear Hall (NLH) effect and is related to the dipole moment of Berry curvature or quantum metric. More recently, another type of nonlinear planar Hall effect was proposed with $j^{\textrm{H}} \propto E^2 B$ where the Hall current $j^{\textrm{H}}$, driving electrical field $E$, and magnetic field $B$ lie within the same plane \cite{He2019Jul, Huang2023Mar} as shown in Fig. \ref{fig1}(b), which we call type-II NLH effect and corresponds to the Berry connection polarizability. In the standard Hall geometry, as shown in Fig. \ref{fig1} (c), the ordinary \textit{linear} Hall effect induced by Lorentz force usually dominates, with the Hall current $j^{\textrm{H}} \propto \tau^2 EB$ (where $\tau$ is the relaxation time), and resultant  ordinary Hall conductivity depends only on the carrier density, independent of the electronic wave function properties \cite{Nagaosa2010May, Tan2021Jun} (see also Appendix \ref{Appen_1}). However, the nonlinear properties of the Lorentz force-induced Hall effect have not been explored so far. This raises a straightforward question: does a Lorentz force-induced NLH effect exist, and could this NLH effect further connect the measurable quantities with any refined intrinsic properties of Bloch electrons, beyond just the carrier density like the ordinary Hall effect?

In this paper we show that the Lorentz force together with the Berry-curvature-induced anomalous velocity \cite{Xiao2010Jul, Karplus1954Sep} can lead to a new type of geometrical NLH effect (type-III) with $j^{\textrm{H}} \propto \tau E^2 B$, which is in lower power of $\tau$ compared to the ordinary Hall effect and is related to the more refined geometrical properties of Bloch wave functions, offering a straightforward method to investigate this more intricate geometrical properties in materials. Different from the type-I NLH effect induced by either Berry curvature dipole (BCD) \cite{Sodemann2015Nov, Du2018Dec} or quantum metric dipole \cite{Gao2014Apr, Wang2021Dec, Wang2023Jul}, the proposed type-III NLH effect here results from the Berry curvature \textit{square} dipole which dominates in topological materials with large Berry curvature and can exist in a broader class of magnetic point groups. Based on effective model analysis and density functional theory (DFT) calculations we find sizable type-III NLH conductivity in gate-tuned MnBi$_2$Te$_4$ double septuble layers (SLs) under moderate uniaxial strain. Moreover, we also find giant unidirectional magnetoresistance (UMR) effect in this material system without strain, which can be utilized to build an electronic transistor device prototype. Finally, a symmetry analysis is carried out to identify all magnetic point groups that allow the existence of our NLH and UMR effects. The unique symmetry properties of Berry curvature square dipole can help us find some systems that allow only type-III NLH while forbid type-I NLH effects mentioned above. Our findings pave the way for further investigation on this new type of nonlinear Hall effect and possible electronic device applications.

\section{Theory and analysis}
\subsection{Theory of type-III NLH effect} To understand the origin of type-III NLH effect, we start from the well-known anomalous Hall current of solids \cite{Nagaosa2010May, Xiao2010Jul, Moore2010Jul}:
\begin{equation}\label{j_H}
    \bs{j}^{\text{H}} = -\frac{e^2}{\hbar} \int \frac{d\bs{k}}{(2\pi)^d}~ \bs{E} \times \bs{\Omega} f,
\end{equation}
where $\bs{E}$ is the external driving electrical field; $\bs{\Omega} = i \langle \bs{\nabla_k} u_{\bs{k}} | \times | \bs{\nabla_k} u_{\bs{k}} \rangle$ is the Berry curvature with $| u_{\bs{k}} \rangle$ being the periodic part of the Bloch wave function; $f$ is the electronic occupation number and $d$ refers to the spatial dimension. According to Eq. \eqref{j_H}, the linear anomalous Hall conductivity is totally determined by the electronic Berry curvature at thermodynamic equilibrium, which is finally determined by the \textit{unperturbed} electronic wave function of occupied states. Thus in order to get nonlinear Hall effects, one must consider the field-induced perturbation effect on either the electronic wave function $|u_{\bs{k}}\rangle$ or the distribution function $f$. The NLH effects corresponding to the former one is an intrinsic material property which can be expressed in terms of Berry connection polarizability and quantum metric tensors \cite{Gao2014Apr, Lai2021Aug, Liu2021Dec, Wang2021Dec, Wang2023Jul, Kaplan2024}.

On the other hand the perturbation on the distribution function $f$ can be determined by Boltzmann equation under relaxation time approximation:
\begin{equation}\label{Boltz}
    - \frac{f - f_0}{\tau} = \dot{\bs{k}} \cdot \bs{\nabla_k} f + \dot{\bs{r}} \cdot \bs{\nabla_r} f + \partial_t f
\end{equation}
with $f_0 = \left[ e^{(\varepsilon_{\bs{k}} - \varepsilon_F)/k_BT} + 1 \right]^{-1}$ being the equilibrium Fermi-Dirac distribution function. Combined with the semiclassical equation of motion of electrons \cite{Xiao2010Jul, Chang1996Mar}, the nonequilibrium contribution to the nonlinear Hall current can be derived as (see details in Appendix \ref{Appen_2}):
\begin{gather}
    \tilde{\bs{j}}^{\text{NLH}} = -\frac{e^2}{\hbar} \int \frac{d\bs{k}}{(2\pi)^d}~ \bs{E} \times \bs{\Omega} \delta f, \label{j_NLH_total} \\
    \delta f= f - f_0 = -\tau \left( \bs{F}_E \cdot \tilde{\bs{v}} + \bs{F}_L \cdot \bs{v}_A \right) \frac{\partial f_0}{\partial \varepsilon_{\bs{k}}} + O(\tau^2). \label{df}
\end{gather}
Here $\bs{F}_E = -e \bs{E}$ is the electrical force, and $\tilde{\bs{v}} = \bs{v}+\bs{v_M}$ is the total group velocity, with $\bs{v_M}= \bs{\nabla_k} (-\bs{m_k}\cdot \bs{B})/ \hbar$, and  $\bs{v} = \bs{\nabla_k} \varepsilon_{\bs{k}}/\hbar$ the ordinary group velocity of electron. The ordinary group velocity part in the first term on the right-hand side of Eq. \eqref{df} corresponds to the Berry curvature dipole contribution to the nonliner anomalous Hall effect \cite{Sodemann2015Nov}, and has been extensively studied in inversion-symmetry-breaking nonmagnetic materials \cite{Kang2019Apr, Ma2019Jan, Du2018Dec}.

The second term, on the other hand, is the Lorentz-force-induced contribution which is seldom discussed before and will be the focus of this paper hereafter. Here, $\bs{F}_L = -e \bs{v} \times \bs{B}$ refers to the Lorentz force which is perpendicular to the group velocity. Therefore, the Lorentz force does not induce energy shift in non-topological materials with zero Berry curvature. However, in topological materials the velocity of electrons acquires an anomalous term $\bs{v}_A = (-e/\hbar) \bs{E} \times \bs{\Omega}$ induced by the Berry curvature effects \cite{Xiao2010Jul}. The $\bs{F}_L \cdot \bs{v}_A \tau$ can be viewed as the energy shift induced by the Lorentz force and anomalous velocity which is generally nonzero and generates the Fermi surface shift shown in Fig. \ref{fig1} (d). It should be noted that the magnetic moment-related part in the first term on the right hand side of Eq. \eqref{df} is of the same order $O(\tau E^2B)$ as the Lorentz force-induced one but exhibits relatively small magnitude \cite{SM}. Therefore, we will focus only on the Lorentz-force part.

Equations \eqref{j_NLH_total} and \eqref{df} demonstrate a new type of NLH effect induced by the combination of Lorentz force and the anomalous velocity. Given that $\bs{F}_L \propto \bs{B}$ and $\bs{v}_A \propto \bs{E}$, the NLH current is proportional to $\tau E^2B$ and it can be expressed in a compact form as $j^{\text{NLH}}_\alpha = \sum_{\beta,\gamma,\lambda} \sigma^{\text{NLH}}_{\alpha\beta\gamma\lambda} E_\beta E_\gamma B_\lambda$ with $\sigma^{\text{NLH}}_{\alpha\beta\gamma\lambda}$ being the NLH conductivity whose expression is given by
\begin{equation}\label{CNLH}
    \sigma^{\text{NLH}}_{\alpha\beta\gamma\lambda} = \frac{e^4\tau}{\hbar^2} \int \frac{d\bs{k}}{(2\pi)^d}~ \sum_\kappa \epsilon_{\alpha\beta\kappa} \Omega_\kappa ( v_\gamma \Omega_\lambda - \delta_{\gamma\lambda} \sum_{\mu}v_\mu \Omega_\mu ) \frac{\partial f_0}{\partial \varepsilon_{\bs{k}}},
\end{equation}
where Greek indexes represents Cartesian coordinates, $\epsilon_{\alpha\beta\kappa}$ is the Levi-Civita tensor, and $\delta_{\gamma\lambda}$ is the Kronecker symbol. Recall that the group velocity $\bs{v}$ is odd under both inversion symmetry ($\mathcal{P}$) and time-reversal symmetry ($\mathcal{T}$) while the Berry curvature $\bs{\Omega}$ is even under $\mathcal{P}$ but odd under $\mathcal{T}$ so $\sigma^{\textrm{NLH}}_{\alpha\beta \gamma\lambda}$ exists only when $\mathcal{P}$ and $\mathcal{T}$ are simultaneously broken.


In addition to the intrinsic contribution from Berry curvature, disorder scatterings, such as skew scattering and side jumps, also play a role in linear Hall conductivity \cite{Sinitsyn2005, Sinitsyn2008Jan, DHou2015}. These scatterings are naturally expected to influence the nonlinear Hall (NLH) conductivity as well \cite{Rhonald2023}. The impact of disorder scattering is highly dependent on carrier density: it is minimal when the Fermi energy is near the band edge but becomes significant as carrier density increases \cite{Du2019Jul}. Conversely, Eq. \eqref{CNLH} demonstrates that the NLH conductivity proposed here is proportional to the square of the Berry curvature on the Fermi surface. This indicates its prominence in small-gap topological materials when the Fermi level is close to the band edges. The stark contrast between the contributions from Berry curvature and disorder scatterings helps to distinguish the mechanisms behind NLH conductivity through electronic doping effects.

\subsection{Berry curvature square dipole} For 2D materials lying within $xy$ plane, in order that the Lorentz force can take effects the magnetic field must have $z$ component. This is reflected by fact that the NLH conductivity $\sigma^{\text{NLH}}_{\alpha\beta\gamma\lambda}$ is nonzero only when $\lambda=z$, which is dramatically different from the nonlinear planar Hall effect \cite{He2019Jul, Huang2023Mar}. Moreover, the second term on the right hand side of Eq. \eqref{CNLH} vanishes for 2D materials because $\bs{v} \perp \bs{\Omega}$. Thus the NLH conductivity can be rewritten as:
\begin{gather}
    \sigma^{\textrm{NLH}}_{yxxz} = -\sigma^{\textrm{NLH}}_{xyxz} = \frac{e^4\tau}{\hbar^3} \int \frac{d\bs{k}}{(2\pi)^2}~ (\partial_{k_x} \Omega^2_z) f_0, \\
    \sigma^{\textrm{NLH}}_{xyyz} = -\sigma^{\textrm{NLH}}_{yxyz} = -\frac{e^4\tau}{\hbar^3} \int \frac{d\bs{k}}{(2\pi)^2}~ (\partial_{k_y} \Omega^2_z) f_0,
\end{gather}
while other components are zero. The above expressions show that the NLH conductivity is proportional to the dipole moment of square of Berry curvature over occupied states, which is defined as
\begin{equation}
    \mathcal{D}^{(n)}_{\alpha} = \int \frac{d\bs{k}}{(2\pi)^2}~ (\partial_{k_\alpha} \Omega^n_z) f_0,
\end{equation}
with $n=2$. Because $\mathcal{D}^{(n)}_\alpha$ behaves like a vector within the 2D plane, the NLH conductivity shows a peculiar angular dependence: for applied electric field $\bs{E} = E(\cos\theta, \sin\theta, 0)$, the NLH conductivity is determined by $\sigma^{\textrm{NLH}}(\theta) = (e^4\tau/\hbar^3) ( \mathcal{D}^{(2)}_x \cos\theta + \mathcal{D}^{(2)}_y \sin\theta )$. It should be noted that $\mathcal{D}^{(2)}_\alpha$ is different from the ordinary BCD \textit{i.e.} $\mathcal{D}^{(1)}_\alpha$ first proposed in Ref. \cite{Sodemann2015Nov}. Under vertical mirror symmetry, $\mathcal{D}^{(1)}_\alpha$ behaves like a \textit{pseudo} vector which should be perpendicular to the mirror plane, while $\mathcal{D}^{(2)}_\alpha$ behaves like a \textit{real} vector and should be parallel to the mirror plane. Moreover, $\mathcal{D}^{(2)}_\alpha$ is expected to become more prominent than $\mathcal{D}^{(1)}_\alpha$ near the band crossings and anticrossings due to the large Berry curvature.

\begin{figure}
\centering
\includegraphics[width=\linewidth]{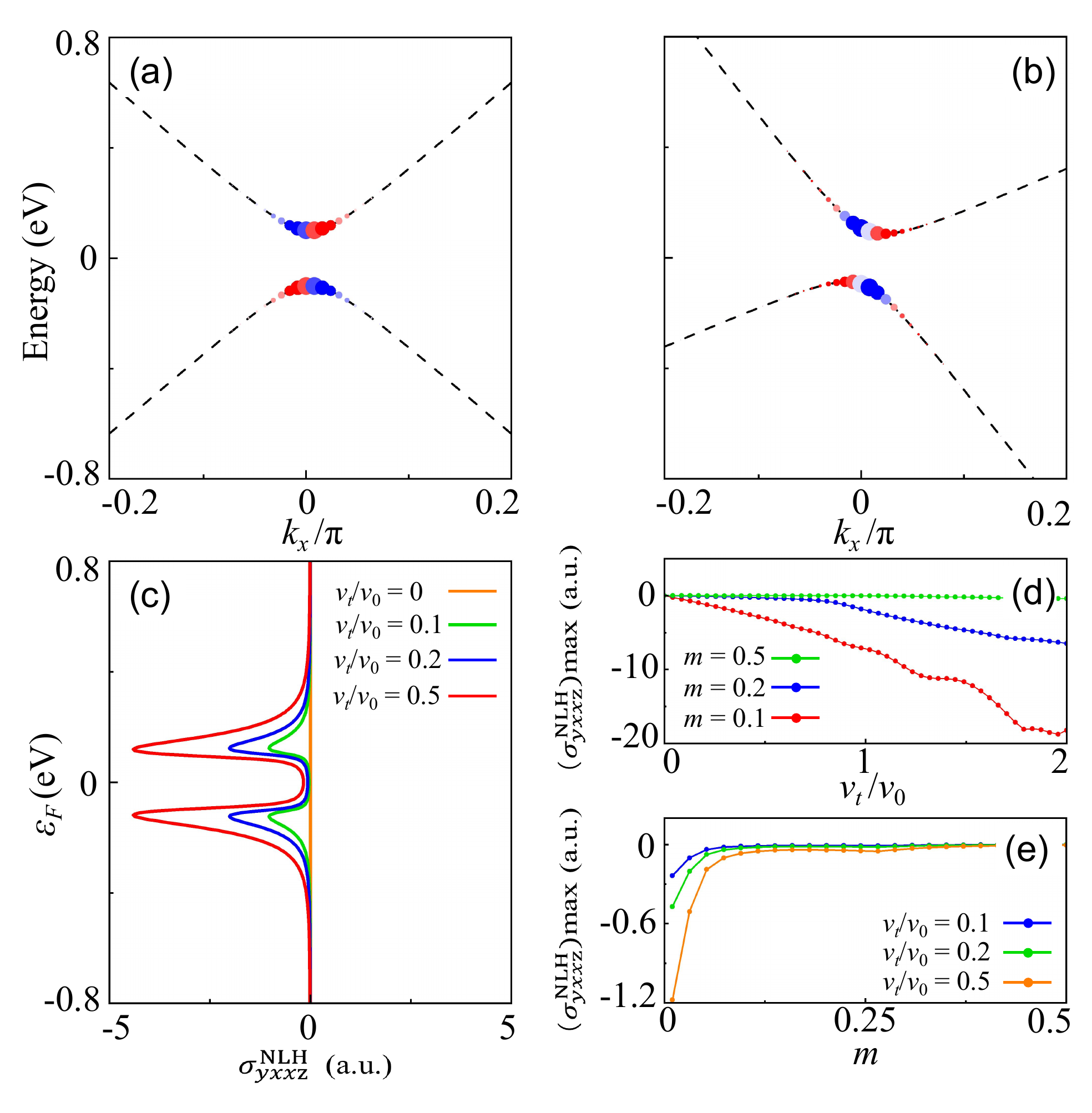}
\caption{\label{fig2}
(a)-(b) Band-resolved $\sigma^{\textrm{NLH}}_{yxxz}$ of the two-band model Eq. \eqref{Dirac} for (a) $v_t=0$ and (b) $v_t=0.5v_0$, respectively. Fixed parameters $\hbar v_0 = 1$ eV $\cdot$ \AA ~ and $m=0.2$ eV are used for both calculations. Red (Blue) circles indicate positive (negative) values. For untilted model the positive and negative values cancel exactly for any Fermi energy $\varepsilon_F$ thus giving zero $\sigma^{\textrm{NLH}}_{yxxz}$, whereas the tilted model can have nonzero $\sigma^{\textrm{NLH}}_{yxxz}$. (c) Calculated $\sigma^{\textrm{NLH}}_{yxxz}$ as function of $\varepsilon_F$ with fixed $m=0.2$ eV but varying $v_t$. The maximum values of $\sigma^{\textrm{NLH}}_{yxxz}$ occur near the band edges where Berry curvature become large. (d)-(e) Calculated maximum $\sigma^{\textrm{NLH}}_{yxxz}$ for varying $v_t$ and $m$. Large $\sigma^{\textrm{NLH}}_{yxxz}$ is favored by large $v_t$ and small $m$.
}
\end{figure}

\section{Results and discussions}
\subsection{Effective Model analysis} We take the 2D massive Dirac model as an example to demonstrate the behavior of NLH conductivity here. Without loss of generality we consider $\sigma^{\textrm{NLH}}_{yxxz}$ here as an example. A nonzero $\sigma^{\textrm{NLH}}_{yxxz}$ requires the breaking of mirror symmetry $M_x$. The minimal effective model can be expressed as,
\begin{equation}\label{Dirac}
    \begin{split}
        H_{\bs{k}}= \hbar v_t k_x + \hbar v_0(k_x\sigma_x + k_y\sigma_y) + m\sigma_z,
    \end{split}
\end{equation}
with $v_0$ being the group velocity, $m$ being the mass term, and $v_t$ being the tilting parameter to break $\mathcal{P}$, $\mathcal{T}$ and $M_x$ symmetries. The band dispersion and Berry curvature are  $\varepsilon_{s\bs{k}} = \hbar v_t k_x + s \sqrt{ \hbar^2 v_0^2 \bs{k}^2+ m^2}$, and $\Omega_z = -s\frac{m\hbar^2 v^2_0}{2(\hbar^2 v^2_0 \bs{k}^2 + m^2)^{3/2}}$ ($s= \pm $ represents the upper or lower band), respectively, with a band gap of $2|m|$. For small tilting, we can derive that $\sigma^{\textrm{NLH}}_{yxxz}(\varepsilon_F) = \frac{e^4\tau m^2 v^2_0 v_t}{8\pi|\varepsilon_F|^5} \left( 1- \frac{3m^2}{|\varepsilon_F|^2} \right) + O(v^2_t)$ with $\left|\varepsilon_F\right| \geq m$ to ensure nonzero density of states \cite{SM}. The maximum $\sigma^{\textrm{NLH}}_{yxxz}$ occurs at the band edges $|\varepsilon_F|=m$, and the value is proportional to $m^{-3}$ which is prominent in small-gap materials.

Figure \ref{fig2} shows the numerical results for the effective model. For untilted model, the band resolved $\sigma^{\textrm{NLH}}_{yxxz}$ exhibits opposite values between $\bs{k}$ and $-\bs{k}$ which gives vanishing $\sigma^{\textrm{NLH}}_{yxxz}$ for any $\varepsilon_F$ so finite tilting is necessary to get nonzero $\sigma^{\textrm{NLH}}_{yxxz}$ [Figs. \ref{fig2}(a)-(b)]. The calculated $\sigma^{\textrm{NLH}}_{yxxz}$ increases with increasing $v_t$ and the maximum values occurs near the band edge [Fig. \ref{fig2}(c)], which is consistent with the analytical result. Figures \ref{fig2}(d)-(e) show the calculated maximum $\sigma^{\textrm{NLH}}_{yxxz}$ with varying $v_t$ and $m$ favoring large tilting and small band gap.

\begin{figure}
\includegraphics[width=\linewidth]{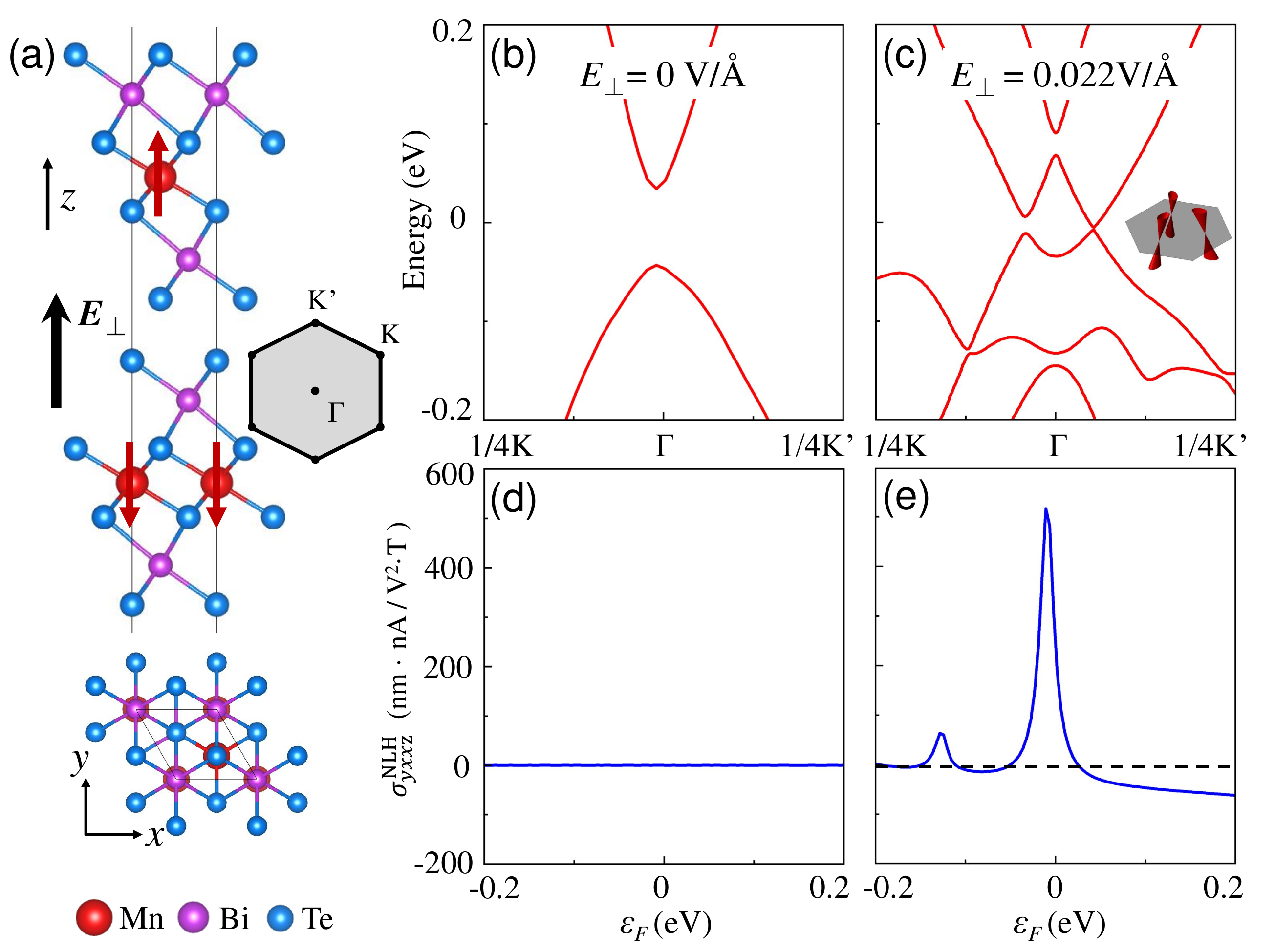}
\caption{
(a) Crystal sturcture and Brillouin zone of MnBi$_2$Te$_4$ double septuble layers (SLs). The red arrows refer to the antiferromagnetic order between adjacent SLs. The crystal and magnetic structure has $\mathcal{PT}$ symmetry. $\bs{E}_\perp$ refers to externally applied vertical electric field which breaks $\mathcal{P}$. (b)-(c) Band structures of MnBi$_2$Te$_4$ double SLs with (b) $E_\perp=0$ and (c) $E_\perp = 0.022$ V/\AA, respectively. The inset of (c) shows the schematic distribution of tilted Dirac cones related by $C_{3z}$ symmetry. (d) Calculated conductivity $\sigma^{\textrm{NLH}}$ as a function a fermi energy $\varepsilon_F$ in strained MnBi$_2$Te$_4$ double SLs with electric field same as that in (b) or in MnBi$_2$Te$_4$ double SLs without strain. (e) Calculated $\sigma^{\textrm{NLH}}$ as a function a fermi energy $\varepsilon_F$ in strained MnBi$_2$Te$_4$ double SLs with electric field same as that in (c). For details of the calculation methods, see \cite{SM}.
\label{fig3}
}
\end{figure}

\subsection{Candidate Material} According to the effective model analysis, $\mathcal{P}$- and $\mathcal{T}$- breaking materials with small band gap (\textit{i.e.}, near the topological phase transition) favor large NLH conductivity. MnBi$_2$Te$_4$ is a recently discovered layered topological antiferromagnet which attracts extensive research of interest because of its unique axion dynamics and versitile topological phase transitions \cite{Gong2019Jun, Li2019Jun, Bernevig2022Mar}. Remarkably, recent studies have shown the gate-tunable topological properties of MnBi$_2$Te$_4$ thin film \cite{Du2020May} with multiple Dirac Fermions near the Fermi energy under suitable vertical electric gating field. Figure \ref{fig3}(a) shows the crystal structure and Brillouin zone of MnBi$_2$Te$_4$ double septuple layers (SLs). A vertical electric field $E_\perp$ breaks the $\mathcal{PT}$ symmetry thus giving rise to finite Berry curvature. Without the vertical electric field, MnBi$_2$Te$_4$ with double SLs is a topologically trivial insulator with spin degeneracy protected by $\mathcal{PT}$ [Fig. \ref{fig3}(b)]. Applying the vertical electric field will split the energy bands and reduce the band gap \cite{SM}
. As increasing $E_\perp$, the band gap closes at about a critical field $E_\perp = E_c =$ 0.022 V/\AA \enspace with a tilted Dirac cone at the Fermi level [Fig. \ref{fig3}(c)]. Due to the 3-fold rotational symmetry $C_{3z}$, there are three tilted Dirac cones whose tilting directions are related by $C_{3z}$. Therefore, the net NLH conductivity vanishes [Fig. \ref{fig3}(d)]. By applying an external uniaxial strain the $C_{3z}$ is broken and a nonzero NLH conductivity arises. Figure \ref{fig3}(e) shows the calculated $\sigma^{\textrm{NLH}}_{yxxz}$ as function of Fermi energy $\varepsilon_F$ with a large peak which favors further experimental measurement.

\begin{figure}
\includegraphics[width=\linewidth]{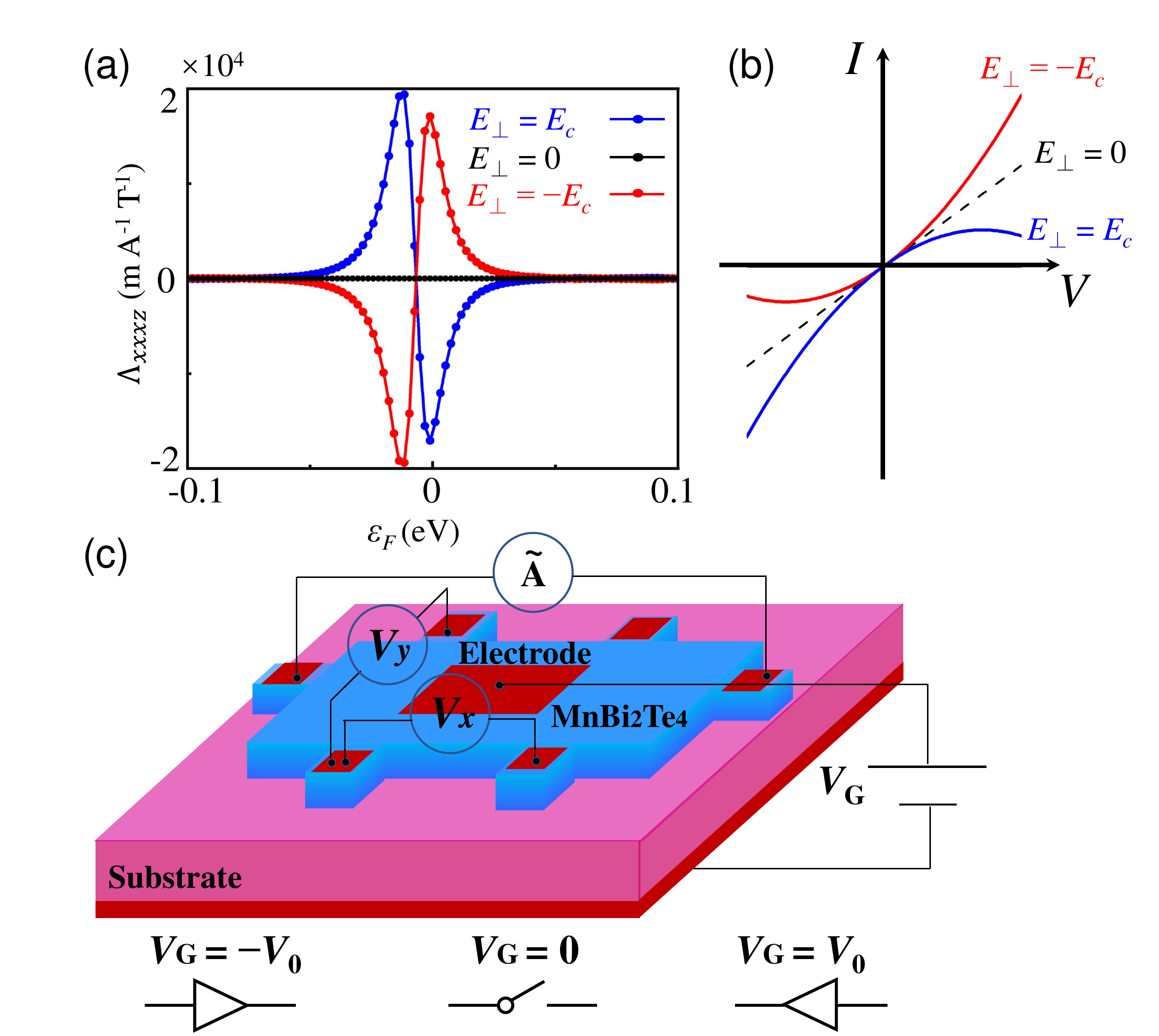}
\caption{ (a) Calculated $\Lambda_{xxxz}$ in gated MnBi$_2$Te$_4$ double SLs with different vertical electirc field $E_\perp = \pm E_c$ or $E_\perp = 0$. (b) Schematic of I-V curve with different values of $E_\perp$. (c) Schematic view of the transistor device prototype, where the critical electric field with positive (negative) value corresponds to the forward (backward) conducting case while the zero electric field corresponds to the insulating case. \label{fig4}}
\end{figure}

\subsection{Unidirectional magnetoresistance} In the presence of external magnetic field, in addition to the NLH effect discussed above, there is another intriguing transport phenomenon that the longitudinal conductivity $\sigma$ changes with the direction of current, which is called electrical magnetrochiral anisotropy or unidirectional magnetoresistance (UMR) effect, and has the same order $O(E^2B)$ with NLH. The UMR effect in gated MnBi$_2$Te$_4$ double SLs can be described by current- and magnetic-field- dependent conductivity $\sigma_{\alpha\beta}(\bs{j},\bs{B}) = \sigma^0_{\alpha\beta} \left( 1 + \sum_{\gamma,\lambda} \Lambda_{\alpha\beta\gamma\lambda} j_\gamma B_\lambda \right)$ with $\sigma^0_{\alpha\beta}$ being the ordinary Drude conductivity and $\Lambda_{\alpha\beta\gamma\lambda}$ being an intrinsic material property. Based on semiclassical equation of motion, the UMR is dominated by the Berry curvature on the Fermi surface, which is confirmed by the analytical results in a 3D Weyl model \cite{SM}. Figure \ref{fig4}(a) shows the calculated $\Lambda_{xxxz}$ under perpendicular electrical field $E_\perp = \pm E_c$. The calculated UMR shows sharp peaks of about $2\times10^4$ m~A$^{-1}$~T$^{-1}$ around $\varepsilon_F = 0$ and its sign reverses with the direction of $E_\perp$, which means the magnitude of electrical conductivity changes dramatically for a sample under current $j \approx 1$~$\mu$A/cm and $B_z=1$ T. Such a nonlinear conductivity can be used to realize a electrical transistor device prototype under a small gating field $V_0$ which induces $E_\perp = \pm E_c$ [Figs. \ref{fig4}(b)-(c)]. Besides, we find that NLH effect can exhibit sizable signals up to moderate temperature, however, UMR effect survives only in the low-teperature regime \cite{SM}.

\begin{table}
\centering
\begin{tabular}{l c c c}
\hline
\hline
    Magnetic point groups  &  NLH  &  BCD & INHE \\
\hline
    $-$6, 6$^{\prime}$/m, $-$6m2, $-$6m$^{\prime}$2$^{\prime}$, 6$^{\prime}$/mmm$^{\prime}$, 23,  &  \Checkmark & \XSolidBrush & \XSolidBrush \\
    m$^{\prime}$$-$3$^{\prime}$, 432, 4$^{\prime}$32$^{\prime}$, $-$43m, $-$4$^{\prime}$3m$^{\prime}$, m$^{\prime}$$-$3$^{\prime}$m, \\
    m$^{\prime}$$-$3$^{\prime}$m$^{\prime}$ \\
\hline
    6$^{\prime}$, 6$^{\prime}$22$^{\prime}$, 6$^{\prime}$mm$^{\prime}$ &  \Checkmark  &  \Checkmark  &  \XSolidBrush \\
\hline
    $-$1$^{\prime}$, 2$^{\prime}$/m, 2/m$^{\prime}$, m$^{\prime}$mm, m$^{\prime}$m$^{\prime}$m$^{\prime}$, 4/m$^{\prime}$, &  \Checkmark  &  \XSolidBrush  &  \Checkmark \\
    4$^{\prime}$/m$^{\prime}$, 4/m$^{\prime}$mm, 4$^{\prime}$/m$^{\prime}$m$^{\prime}$m, 4/m$^{\prime}$m$^{\prime}$m$^{\prime}$, \\
    $-$3$^{\prime}$, $-$3$^{\prime}$m, $-$3$^{\prime}$m$^{\prime}$, $-$6$^{\prime}$, 6/m$^{\prime}$, $-$6$^{\prime}$m$^{\prime}$2, \\
    $-$6$^{\prime}$m2$^{\prime}$, 6/m$^{\prime}$mm, 6/m$^{\prime}$m$^{\prime}$m$^{\prime}$ \\
\hline
    11$^{\prime}$, 21$^{\prime}$, m1$^{\prime}$, 2221$^{\prime}$, mm21$^{\prime}$, 41$^{\prime}$, $-$41$^{\prime}$,  &  \XSolidBrush  &  \Checkmark  &  \XSolidBrush \\
    4221$^{\prime}$, 4mm1$^{\prime}$, $-$42m1$^{\prime}$, 31$^{\prime}$, 321$^{\prime}$, \\
    3m1$^{\prime}$, 61$^{\prime}$, 6221$^{\prime}$, 6mm1$^{\prime}$ \\
\hline
    1, 2, 2$^{\prime}$, m, m$^{\prime}$, 222, 2$^{\prime}$2$^{\prime}$2, mm2, m$^{\prime}$m2$^{\prime}$, &  \Checkmark  &  \Checkmark  &  \Checkmark \\
    m$^{\prime}$m$^{\prime}$2, 4, 4$^{\prime}$, $-$4, $-$4$^{\prime}$, 422, 4$^{\prime}$22$^{\prime}$, 42$^{\prime}$2$^{\prime}$, \\
    4mm, 4$^{\prime}$m$^{\prime}$m, 4m$^{\prime}$m$^{\prime}$, $-$42m, $-$4$^{\prime}$2$^{\prime}$m,  \\
    $-$4$^{\prime}$2m$^{\prime}$, $-$42$^{\prime}$m$^{\prime}$, 3, 32, 32$^{\prime}$, 3m, 3m$^{\prime}$, 6, \\
    622, 62$^{\prime}$2$^{\prime}$, 6mm, 6m$^{\prime}$m$^{\prime}$ \\
\hline
\hline
\end{tabular}
    \caption{Table of magnetic point groups classified by the existence or absence of Lorentz force induced NLH, BCD induced NLAH, and intrinsic nonlinear Hall effect (INHE), where \Checkmark means at least one component of the tensor is nonzero.}
    \label{TABLE 1}
\end{table}

\subsection{Symmetry analysis} Now we analyze the symmetry properties of NLH and UMR effects. The UMR generally requires inversion symmetry breaking and can exist in all non-centrosymmetric magnetic point groups. On the other hand, the NLH effect is more complicated. Type-III NLH effect has a different physical origin from the BCD induced NLAH and the intrinsic nonlinear Hall effect (INHE), both of which belong to the type-I NLH effect, thus has different symmetry requirements. By careful analysis, we give the list of magnetic point groups that allow the existence or absence of type-III NLH effect, along with that of BCD and INHE, as shown in Table \ref{TABLE 1} which is obtained based on the magnetic tensor symmetry module implemented in Bilbao Crystallographic Server \cite{Elcoro2019May}. It should be noted that BCD is fully forbidden by the coexistence of a mirror symmetry (which forbids the diagonal parts of BCD) and a rotation symmetry (which forbids the off-diagonal parts of BCD), and the case is similar for INHE which origins from quantum metric dipole, while type-III NLH effect is related to the Berry curvature square dipole which can at least have nonzero diagonal elements in the presence of a mirror symmetry. Obviously, we can find some magnetic groups where our type-III NLH exists but BCD induced NLAH and INHE are forbidden (the first row of Table \ref{TABLE 1}). There are also some magnetic groups that allow BCD but forbid type-III NLH effect, which all have time reversal symmetry (the fourth row of Table \ref{TABLE 1}). Additionally, from Table \ref{TABLE 1}, we can clearly see that the type-III NLH effect can exist in a wider range among all the magnetic point groups (69 of 122 magnetic point groups allow type-III NLH) than BCD and INHE (53 of 122 magnetic point groups allow BCD and INHE). This broader range may benefit further experimental investigation and verification.

\section{Summary} We propose a theory of a new type of nonlinear Hall effect, which stems from the combination of the Lorentz force and anomalous velocity. Based on first-principles calculations, we show MnBi$_2$Te$_4$ double SLs as an ideal candidate with significant NLH conductivity. Interestingly, a giant UMR effect is also predicted in this material system with full electrical tunability, which may be useful for developing a new generation of high-performance electrically switchable transistor without PN junction. Our symmetry analysis further shows that this NLH effect can exist in a wider range of magnetic point groups compared to previously reported mechanisms, which may facilitate further investigations into these effects. Our finding highlights the important role of the Lorentz force in exploring the more refined electronic structure and topology of materials, which may have been previously overlooked.

\section{Acknowledgements}
This work was supported by the Innovation Program for Quantum Science and Technology (Grant No. 2023ZD0300500), the Basic Science Center Project of NSFC (Grant No. 51788104), and the Beijing Advanced Innovation Center for Future Chip. Y.L. is sponsored by the Shanghai Pujiang Program (Grant No. 23PJ1413000), NSFC (Grant No. 12404279) and the Fundamental Research Funds for the Central Universities.


\appendix

\section{Derivation of ordinary Hall (OH) conductivity \label{Appen_1}}

In this section we derive the expression of ordinary Hall (OH) conductivity for a nearly free electronic gas model that supports a parabolic band. For a nearly free electron gas (without Berry curvature or orbital magnetic moment), the semiclassical equation of motion is
\begin{equation}
    \begin{split}\label{EOM_0}
        \hbar \dot{\bs{k}} =& -e\bs{E} -e\dot{\bs{r}} \times \bs{B}, \\
        \dot{\bs{r}} =& \bs{v} = \frac{1}{\hbar} \bs{\nabla_k} \varepsilon_{\bs{k}}.
    \end{split}
\end{equation}
The Boltzmann transport equation is
\begin{equation*}
    \left( \partial_t + \dot{\bs{k}} \cdot \bs{\nabla_k} + \dot{\bs{r}} \cdot \bs{\nabla_r} \right) f = - \frac{f - f_0}{\tau},
\end{equation*}
\textit{i.e.},
\begin{equation}\label{Boltz_eq2}
    \left[ 1 + \tau\left( \partial_t + \dot{\bs{k}}\cdot \bs{\nabla_k} + \dot{\bs{r}}\cdot \bs{\nabla_r} \right) \right] f = f_0.
\end{equation}
Because $\tau$ is usually small, the above Eq. \eqref{Boltz_eq2} can be formally solved as
\begin{equation}
    \begin{split}
        f =& \left[ 1 + \tau \left( \partial_t + \dot{\bs{k}}\cdot \bs{\nabla_k} + \dot{\bs{r}}\cdot \bs{\nabla_r} \right) \right]^{-1} f_0 \\
        =& \sum^{+\infty}_{l=0} \left[ - \tau \left( \partial_t + \dot{\bs{k}}\cdot \bs{\nabla_k} + \dot{\bs{r}}\cdot \bs{\nabla_r} \right) \right]^l f_0 \\
        =& \sum^{+\infty}_{l=0} f_l,
    \end{split}
\end{equation}
where $f_l \propto \tau^l$ is the $l$-th order perturbation term for the distribution function. For the steady and uniform distribution, \textit{i.e.}, $\partial_t f = \bs{\nabla_r} f =0$, the expression of $f_l$ can be simplified as
\begin{equation}\label{fl}
    f_l = \left( -\tau \dot{\bs{k}} \cdot \bs{\nabla_k} \right)^l f_0.
\end{equation}

Now we can derive the expression of OH conductivity based on Eqs. \eqref{EOM_0} and \eqref{fl}. Substitute Eq. \eqref{EOM_0} into \eqref{fl}, we get the first order term as
\begin{equation}\label{f1_ord}
    \begin{split}
        f_1 =& \frac{e\tau}{\hbar} ( \bs{E} + \bs{v} \times \bs{B}) \cdot \bs{\nabla_k} f_0 \\
        =& \frac{e\tau}{\hbar} ( \bs{E} + \bs{v} \times \bs{B} ) \cdot \hbar \bs{v} \frac{\partial f_0}{\partial \varepsilon_{\bs{k}}} \\
        =& e\tau \bs{E} \cdot \bs{v} \frac{\partial f_0}{\partial \varepsilon_{\bs{k}}}.
    \end{split}
\end{equation}
From Eq. \eqref{f1_ord}, we can see that Lorentz force $\bs{F}_L =- e\bs{v} \times \bs{B}$ does not enter into the first-order term because it is perpendicular to the group velocity $\bs{v}$. As a result, it does not induce a Fermi surface shift. The second order term is
\begin{equation}
    \begin{split}\label{f2_ord}
        f_2 =& -\tau \dot{\bs{k}} \cdot \bs{\nabla_k} f_1 \\
        =& \frac{e\tau}{\hbar} (\bs{E} + \bs{v} \times \bs{B}) \cdot \bs{\nabla_k} \left[ e \tau \bs{E} \cdot \bs{v} \frac{\partial f_0}{\partial \varepsilon_{\bs{k}}} \right] \\
        =& \frac{e^2\tau^2}{\hbar} ( \bs{E} + \bs{v} \times \bs{B} ) \\
        & \hspace{1cm} \cdot \left[ \bs{\nabla_k} (\bs{v} \cdot \bs{E}) \frac{\partial f_0}{\partial \varepsilon_{\bs{k}}} + (\bs{E} \cdot \bs{v}) \frac{\partial^2 f_0}{\partial \varepsilon^2_{\bs{k}}} \hbar \bs{v} \right].
    \end{split}
\end{equation}
We focus on the Lorentz-force induced terms in Eq. \eqref{f2_ord} which will lead to the OH conductivity:
\begin{equation}
\begin{split}
    f_{\textrm{OH}}&= -\tau^2 \frac{e^2}{\hbar} (\bs{v} \times \bs{B}) \cdot \bs{\nabla_k} (\bs{v} \cdot \bs{E}) \frac{\partial f_0}{\partial \varepsilon_{\bs{k}}} \\
    &=\tau^2 \frac{e^2}{\hbar} [ \bs{\nabla_k} (\bs{v} \cdot \bs{E}) \times \bs{B}] \cdot \bs{v} \frac{\partial f_0}{\partial \varepsilon_{\bs{k}}} \\
    &= \tau^2 e^2 \bs{v} \cdot \left[ \frac{1}{m^*} \bs{E} \times \bs{B} \right] \frac{\partial f_0}{\partial \varepsilon_{\bs{k}}},
\end{split}
\end{equation}
where the dispersion relation of nearly free electron gas $\varepsilon_{\bs{k}} = \frac{\hbar^2 \bs{k}^2}{2m^*}$ and the inverse effective mass $\left(\frac{1}{m^*} \right)_{\alpha\beta} = \frac{1}{\hbar^2} \frac{\partial \varepsilon_{\bs{k}}}{\partial k_\alpha k_\beta}$ have been used. Thus the ordinary Hall current is expressed by
\begin{equation}
    \begin{split}
        \bs{j}^{\textrm{OH}} =& -\int \frac{d\bs{k}}{(2\pi)^d}~ e\bs{v} f_{\textrm{OH}} \\
        =& - \frac{e^3\tau^2}{m^*} \int \frac{d\bs{k}}{(2\pi)^d}~ \bs{v}\bs{v} \cdot \left( \bs{E} \times \bs{B} \right) \frac{\partial f_0}{\partial \epsilon_k}.
    \end{split}
\end{equation}
For isotropic 3D electron gas, the OH effect can be expressed as $\bs{j}^{\textrm{OH}} = \sigma^{\textrm{OH}} (\bs{E} \times \bs{B})$ with the ordinary Hall conductivity calculated as
\begin{equation}
    \begin{split}\label{sigma_OH}
        \sigma^{\textrm{OH}} =& -\frac{e^3\tau^2}{m^*} \int \frac{d\bs{k}}{(2\pi)^3}~ \frac{\bs{v}^2}{3} \frac{\partial f_0}{\partial \varepsilon_{\bs{k}}} \\
        \approx& \frac{e^3 \tau^2}{3m^* (2\pi)^3} \int d\bs{k}~ \frac{\hbar^2 \bs{k}^2}{(m^*)^2} \delta\left(\varepsilon_F - \frac{\hbar^2 \bs{k}^2}{2m^*} \right) \\
        =& \frac{2e^3\tau^2}{3(m^*)^2 (2\pi)^3} \int^{+\infty}_0 dk~ 4\pi k^2 \frac{k^2\delta(k_F - k)}{2k_F} \\
        =& \frac{4\pi e^3 \tau^2 }{3(m^*)^2 (2\pi)^3} k^3_F,
    \end{split}
\end{equation}
where $k_F = \sqrt{2m^* \varepsilon_F}/\hbar$ is the Fermi wave vector, and $\frac{\partial f_0}{\partial \varepsilon_{\bs{k}}} \approx - \delta(\varepsilon_F - \varepsilon_{\bs{k}})$ has been used in the above derivation. On the other hand, the carrier density is
\begin{equation}
    \begin{split}\label{nF}
        n_F =& \int \frac{d\bs{k}}{(2\pi)^3} f_0 = \int^{k_F}_0 dk \frac{4\pi k^2}{(2\pi)^3} = \frac{k^3_F}{6\pi^2}.
    \end{split}
\end{equation}
By substitute Eq. \eqref{nF} into \eqref{sigma_OH}, we get the familiar expression of OH conductivity as
\begin{equation}
    \sigma^{\textrm{OH}} = \frac{e^3\tau^2n_F}{(m^*)^2}.
\end{equation}
Two key points about the ordinary Hall effect:
\begin{itemize}
    \item It is of second order in $\tau$;
    \item It relies only on the band dispersion relation $\varepsilon_{\bs{k}}$ and is independent of the Bloch wave function $|u_{\bs{k}}\rangle$.
\end{itemize}

\section{Derivation of type-III nonlinear Hall (NLH) effect \label{Appen_2}}

In this section, we derive the expression of NLH effect based on semiclassical equation of motion of electrons including Berry curvature and orbital magnetic moments:
\begin{equation}
    \begin{split}\label{s1}
        \hbar\dot{\bs{k}} =& -e \bs{E} - e \dot{\bs{r}} \times \bs{B}, \\
        \dot{\bs{r}} =& \frac{1}{\hbar} \bs{\nabla_k} \varepsilon_M - \dot{\bs{k}} \times \bs{\Omega},
    \end{split}
\end{equation}
where $\varepsilon_M = \varepsilon_{\bs{k}} - \bs{m} \cdot \bs{B}$ ($\bs{m}$ refers to the orbital magnetic moment), and $\bs{\Omega}$ is Berry curvature. Equation \eqref{s1} can be solved as:
\begin{align}
    D \dot{\bs{k}} = -\frac{e}{\hbar} \bs{E} - \frac{e}{\hbar} \tilde{\bs{v}} \times \bs{B} - \frac{e^2}{\hbar^2} (\bs{B} \cdot \bs{E}) \bs{\Omega}, \label{k_dot} \\
    D \dot{\bs{r}} = \tilde{\bs{v}} + \frac{e}{\hbar} \bs{E} \times \bs{\Omega} + \frac{e}{\hbar} (\tilde{\bs{v}} \cdot \bs{\Omega}) \bs{B}, \label{r_dot}
\end{align}
with $\tilde{\bs{v}} = \frac{1}{\hbar} \bs{\nabla_k} \varepsilon_M$ being the modified group velocity, and $D = \left( 1 + \frac{e}{\hbar} \bs{\Omega} \cdot \bs{B} \right)^{-1}$.

We expand the distribution function $f$ in a power series of relaxation time $\tau$ as: $f = f_0 + f_1 + f_2 + \cdots$ with $f_n \propto \tau^n$ and $f_0 = \left[ e^{(\varepsilon_{\bs{k}} - \varepsilon_F)/k_BT} + 1 \right]^{-1}$ being the equilibrium distribution function. According to the Boltzmann equation, i.e., Eq. \eqref{Boltz} of the main text, we have
\begin{equation}\label{f_n}
    \begin{split}
        f_n =& - \tau \dot{\bs{k}} \cdot \nabla_{\bs{k}} f_{n-1},
    \end{split}
\end{equation}
for spatially uniform $\bs{E}$ and $\bs{B}$. Substituting Eq. \eqref{k_dot} into Eq. \eqref{f_n} we can derive the first-order term $f_1$ as:
\begin{equation}\label{f1}
    \begin{split}
        f_1 &= \frac{\tau}{D} \left[ \frac{e}{\hbar} \bs{E} + \frac{e}{\hbar} \tilde{\bs{v}} \times \bs{B} + \frac{e^2}{\hbar^2} (\bs{B} \cdot \bs{E}) \bs{\Omega} \right] \cdot \bs{\nabla_k} f_0 \\
        &= \frac{\tau}{D} \left[ \frac{e}{\hbar} \bs{E} + \frac{e}{\hbar} \tilde{\bs{v}} \times \bs{B} + \frac{e^2}{\hbar^2} (\bs{B} \cdot \bs{E}) \bs{\Omega} \right] \cdot (\bs{\nabla_k} \varepsilon_M) \frac{\partial f_0}{\partial \varepsilon_M} \\
        &\approx \frac{\tau}{D} \left[ e\bs{E} \cdot \tilde{\bs{v}} + e \tilde{\bs{v}} \times \bs{B} \cdot \tilde{\bs{v}} + \frac{e^2}{\hbar} (\bs{B} \cdot \bs{E}) (\bs{\Omega} \cdot \tilde{\bs{v}}) \right] \frac{\partial f_0}{\partial \varepsilon_{\bs{k}}} \\
        &\approx \frac{\tau}{D} \left[ e\bs{E} \cdot \tilde{\bs{v}} + \frac{e^2}{\hbar} (\bs{B} \cdot \bs{E}) (\bs{\Omega} \cdot \bs{v})  \right] \frac{\partial f_0}{\partial \varepsilon_{\bs{k}}},
    \end{split}
\end{equation}
where $\bs{v} = \frac{1}{\hbar} \bs{\nabla_k} \varepsilon_{\bs{k}}$ is the group velocity. On the third line of Eq. \eqref{f1} we have used the approximation $\frac{\partial f_0}{\partial \varepsilon_M} \approx \frac{\partial f_0}{\partial \varepsilon_{\bs{k}}}$ and on the last line we have substituted some $\tilde{\bs{v}}$ terms by $\bs{v_k}$ by omitting a $O(\bs{B}^2)$ term in $f_1$. For small magnetic field $D \approx 1 - \frac{e}{\hbar} \bs{\Omega} \cdot \bs{B}$. Thus the expression of $f_1$ up to linear order of $\bs{B}$ can be simplified as
\begin{equation}
    \begin{split}
        f_1 =& \tau \left[ e\bs{E} \cdot \tilde{\bs{v}} \left( 1 - \frac{e}{\hbar} \bs{\Omega} \cdot \bs{B} \right) + \frac{e^2}{\hbar} (\bs{B} \cdot \bs{E}) (\bs{\Omega} \cdot \bs{v}) \right] \frac{\partial f_0}{\partial \varepsilon_{\bs{k}}} \\
        =& \tau \left[ e\bs{E} \cdot \tilde{\bs{v}} - \frac{e^2}{\hbar} (\bs{E} \cdot \bs{v}) (\bs{\Omega} \cdot \bs{B}) + \frac{e^2}{\hbar} (\bs{B} \cdot \bs{E}) (\bs{\Omega} \cdot \bs{v}) \right] \frac{\partial f_0}{\partial \varepsilon_{\bs{k}}} \\
        =& \tau \left[ e\bs{E} \cdot \tilde{\bs{v}} - \frac{e^2}{\hbar} (\bs{v} \times \bs{B}) \cdot (\bs{E} \times \bs{\Omega}) \right] \frac{\partial f_0}{\partial \varepsilon_{\bs{k}}} \\
        =& -\tau \left( \bs{F}_E \cdot \tilde{\bs{v}} + \bs{F}_L \cdot \bs{v}_A \right) \frac{\partial f_0}{\partial \varepsilon_{\bs{k}}},
    \end{split}
\end{equation}
with $\bs{F}_E = -e\bs{E}$, $\bs{v}_A = (-e/\hbar) \bs{E} \times \bs{\Omega}$, and $\bs{F}_L = -e \bs{v} \times \bs{B}$.


%

\end{document}